\documentclass[11pt,a4paper]{article}

\usepackage{epsfig,amsmath,amssymb}

\begin{document}

\title{Cylindrical thin-shell wormholes} 
\author{Ernesto F. Eiroa$^{1,}$\thanks{e-mail: eiroa@iafe.uba.ar}, 
Claudio Simeone$^{1, 2,}$\thanks{e-mail: csimeone@df.uba.ar}\\
{\small $^1$ Instituto de Astronom\'{\i}a y F\'{\i}sica del Espacio, C.C. 67, 
Suc. 28, 1428, Buenos Aires, Argentina}\\
{\small $^2$ Departamento de F\'{\i}sica, Facultad de Ciencias Exactas y 
Naturales,} \\ 
{\small Universidad de Buenos Aires, Ciudad Universitaria Pab. I, 1428, 
Buenos Aires, Argentina}} 

\maketitle

\begin{abstract}
A general formalism for the dynamics of non rotating cylindrical thin-shell 
wormholes is developed. The time evolution of the throat is
explicitly  obtained for thin-shell wormholes whose metric has the form
associated to local cosmic strings. It is found that the throat collapses to 
zero radius, remains static or expands forever, depending only on the sign 
of its initial velocity.\\ 

\noindent 
PACS number(s): 04.20.Gz, 11.27.+d, 04.40.Nr

\end{abstract}

\section{Introduction}

Traversable Lorentzian wormholes, first studied by Morris and Thorne
\cite{motho}, are solutions of the Einstein field equations that have two 
regions connected by a throat. These \textit{tunnels} can join two parts of 
the same universe or two separate universes \cite{motho, visser}. For static 
wormholes, the throat is defined as a two-dimensional hypersurface of minimal 
area that should satisfy a \textit{flare-out} condition \cite{ hovis1}. For 
time dependent wormholes the general definition of the throat is more complex 
(the interested reader is referred to \cite{hovis2}). All traversable 
wormholes must be threaded by \textit{exotic} matter that violates the null 
energy condition \cite{motho, visser, hovis1, hovis2}. Recently, Visser 
\textit{et al} \cite{viskardad} showed that the amount of \textit{exotic} 
matter that must be present around the throat can be made infinitesimally 
small by a suitable choice of the geometry of the wormhole.\\

In a gauge theory, spontaneous symmetry breaking of a complex scalar field  
leads to cylindrical topological defects known as local cosmic strings 
\cite{vilenkin1}. The gravitational effects of such objects have been 
the object of thorough analysis, because of the possible important 
consequences they could have had for galaxy formation, and also in the study of
gravitational lensing. The spacetime metric around a local cosmic string was 
first obtained  by Vilenkin \cite{vilenkin2} in the linear approximation of 
general relativity. Local strings are characterized by having an 
energy-momentum tensor whose only non null components are 
$T_t^{\, t}=T_z^{\, z}$. Within this framework a Dirac's delta was used to 
model the radial distribution of the energy-momentum tensor for a string along 
the $z$ axis. The resulting  spacetime metric is flat but with a deficit angle 
$\Delta \varphi =8\pi G\mu$, up to first order in $G\mu$ (in GUT strings 
$G\mu \sim 10^{-6}$), with  $\mu$  the linear energy density. Later, 
independently, Gott \cite{gott} and Hiscock \cite{hiscock}, extending 
the analysis to the framework of theories leading to  values of $G\mu$ closer 
to one, showed that the deficit angle found by Vilenkin is actually right to 
all orders in $G\mu$. In their demonstration, they considered a source 
in the form of a thick  cylinder of constant radius with  both uniform 
linear mass density and  tension, the last one only along the axis.
The solution of the full Einstein equations in the interior was matched with  
the vacuum solution for the exterior, and the integration constants appearing 
were determined by matching both metrics in the boundary. In all these 
works, the source determined an exterior solution with Lorentz 
invariance along the $z$ axis. However, if this is not required, 
the most general solution has the well-known Weyl form; which includes two 
solutions with Lorentz invariance, one of them corresponding to the case 
solved by Gott and Hiscock. Later, similar analysis were extended to scalar 
tensor theories of gravity, like that of Brans and Dicke \cite{bd}.
It has also been shown \cite{witten} that a cosmic string can behave as a 
superconductor, with a current along it and a magnetic field in its exterior. 
This can be the result of the appearance of both bosonic or fermionic charge 
carriers. The spacetime around a superconducting string is then not a vacuum 
solution of the Einstein equations, because, besides the effective mass 
associated with  the charge carriers, in this case there is a non localized 
magnetic contribution to the energy-momentum tensor \cite{bapi}.\\

Solutions of the Einstein field equations representing wormholes associated 
with cosmic strings have been previously considered in the literature. 
Cl\'{e}ment \cite{clem1} found traversable multi-wormhole solutions where the 
spacetime metric was asymptotic to the conical cosmic string metric. In other 
work, Cl\'{e}ment \cite{clem2} extended cylindrical multi-cosmic strings 
metrics to wormhole spacetimes with only one region at spatial infinite, and 
analyzed in detail the geometry of asymptotically flat wormhole spacetimes 
produced by one or two cosmic strings. Aros and Zamorano \cite{arza} 
constructed a solution that can be interpreted as a traversable cylindrical 
wormhole inside the core of a global cosmic string.\\

Thin-shell wormholes are made by cutting and pasting two manifolds 
\cite{visser, mvis} to form a geodesically complete new one with a 
throat placed in the joining shell. In this case, the \textit{exotic} 
matter needed to build the wormhole is concentrated at the shell and the 
junction-condition formalism is used for its study.   
Poisson and Visser \cite{poisson} made a linearized stability analysis under 
spherically symmetric perturbations of a thin-shell wormhole constructed by 
joining two Schwarzschild geometries. Later, Barcel\'{o} and Visser 
\cite{barcelo} applied this method to study wormholes constructed using branes 
with negative tensions and Ishak and Lake \cite{ishak} analyzed the stability 
of transparent spherically symmetric thin-shells and wormholes.
Recently, Eiroa and Romero \cite{eirom} extended the linearized stability 
analysis to Reisner-Nordstr\"{o}m thin-shell geometries, and Lobo and Crawford 
\cite{lobo} to wormholes with a cosmological constant.\\

In this article we study cylindrical thin-shell wormholes. We concentrate on 
the geometry of these objects and we do not intend to 
supply any explanation about the mechanisms that might provide the 
\textit{exotic} matter to them. In Sec. 
\ref{cylind} we present the general formalism. In Sec. \ref{vacuum} and 
\ref{supercond}, we apply it to vacuum and 
superconducting cosmic string wormholes. Finally, in Sec. \ref{discu}, 
the results are discussed. Throughout the paper we use units such as $c=G=1$.

\section{Cylindrical thin-shell wormholes}\label{cylind}

The static cylindrically symmetric metric in coordinates $X^{\alpha}=
(t,r,\varphi ,z)$ can be written in the form \cite{thorne}
\begin{equation}
ds^2 = f(r)(-dt^2 +dr^2 )+g(r)d\varphi ^2+h(r)dz^2,
\label{e1}
\end{equation}
where $f$, $g$ and $h$ are positive functions of $r$. From this geometry we
can take two copies\footnote{It is not necessary to take both regions equal, 
but it is enough for our purposes.} of the region with $r \geq a$:
\begin{equation}
  \mathcal{M}^{\pm} = \{ x / r \geq a \}, \label{e2}
\end{equation}
and glue them together at the hypersurface
\begin{equation}
  \Sigma \equiv \Sigma^{\pm} = \{ x / r - a = 0 \}, \label{e3}
\end{equation}
to make a geodesically complete manifold $\mathcal{M}=\mathcal{M}^{+} \cup 
\mathcal{M}^{-}$. If $g_{\varphi \varphi }=g(r)$ is an increasing function for 
$r\in [a,a+\epsilon ]$, with $\epsilon >0$, this construction 
creates a cylindrically symmetric thin-shell wormhole with two regions 
connected by a throat at $\Sigma $. On $\mathcal{M} $ we can define
a new radial coordinate $l = \pm \int_a^r g_{rr} dr$, where the positive and
negative signs correspond, respectively, to $\mathcal{M}^+$ and
$\mathcal{M}^-$, with $|l|$ representing the proper radial distance to the 
throat, which is placed in $l = 0$. To study this traversable wormhole we use 
the standard Darmois-Israel formalism \cite{daris}. For a recent review of
this technique, also called junction-condition formalism, see 
Ref. \cite{mus}.\\

The throat of the wormhole is placed at the shell $\Sigma$, which is a 
synchronous timelike hypersurface. We can adopt coordinates $\xi ^i=(\tau ,
\varphi,z )$ in $\Sigma $, with $\tau $ the proper time on the shell. In order 
to analyze the dynamical behavior, we let the radius of the throat be a 
function of the proper time, $a = a ( \tau )$. Then $\Sigma $ is defined by 
the equation
\begin{equation}
  \Sigma : \mathcal{H} ( r, \tau ) = r - a ( \tau ) = 0. \label{e5}
\end{equation}
The extrinsic curvature (second fundamental forms) associated with the two
sides of the shell are:
\begin{equation}
  K_{ij}^{\pm} = - n_{\gamma}^{\pm} \left. \left( \frac{\partial^2
  X^{\gamma}}{\partial \xi^i \partial \xi^j} + \Gamma_{\alpha \beta}^{\gamma}
  \frac{\partial X^{\alpha}}{\partial \xi^i} \frac{\partial
  X^{\beta}}{\partial \xi^j} \right) \right|_{\Sigma}, \label{e6}
\end{equation}
where $n_{\gamma}^{\pm}$ are the unit normals ($n^{\gamma} n_{\gamma} = 1$) to
$\Sigma$ in $\mathcal{M}$:
\begin{equation}
  n_{\gamma}^{\pm} = \pm \left| g^{\alpha \beta} \frac{\partial
  \mathcal{H}}{\partial X^{\alpha}} \frac{\partial \mathcal{H}}{\partial
  X^{\beta}} \right|^{- 1 / 2} \frac{\partial \mathcal{H}}{\partial
  X^{\gamma}} . \label{e7}
\end{equation}
In the orthonormal basis $\{ e_{\hat{\tau}}, e_{\hat{\varphi}}, e_{\hat{z}}
\}$ ($e_{\hat{\tau}} = \sqrt{1/f(r)}e_{\tau}$, $e_{\hat{\varphi}} 
= \sqrt{1/g(r)}e_{\varphi}$, $e_{\hat{z}} =\sqrt{1/h(r)}e_{z}$, 
$g_{_{\hat{\imath} \hat{\jmath}}} = \eta_{_{\hat{\imath}
\hat{\jmath}}}=diag(-1,1,1,1)$) we have
\begin{equation}
K_{\hat{\tau} \hat{\tau}}^{\pm} = \mp \frac{2f(a)^2 \ddot{a}+f'(a)
+2f'(a)f(a)\dot{a}^2}{2f(a) \sqrt{f(a)} \sqrt{1+f(a)\dot{a}^2}},
\label{e8a}
\end{equation}
\begin{equation}
K_{\hat{\varphi} \hat{\varphi}}^{\pm} = \pm \frac{g'(a)\sqrt{1+f(a)
\dot{a}^2}}{2g(a)\sqrt{f(a)}}, 
\label{e8b}
\end{equation}
and
\begin{equation}
K_{\hat{z} \hat{z}}^{\pm} = \pm \frac{h'(a)\sqrt{1+f(a)\dot{a}^2}}
{2h(a) \sqrt{f(a)}}, 
\label{e8c}
\end{equation}
where the dot means $d / d \tau$.\\

The Einstein equations on the shell reduce to the Lanczos equations:
\begin{equation}
-[K_{\hat{\imath} \hat{\jmath}}]+[K]g_{\hat{\imath} \hat{\jmath}}=
8\pi S_{\hat{\imath} \hat{\jmath}},
\label{e10}
\end{equation}
where $[K_{_{\hat{\imath} \hat{\jmath}}}]\equiv K_{_{\hat{\imath}
\hat{\jmath}}}^+ - K_{_{\hat{\imath} \hat{\jmath}}}^-$, 
$[K]=g^{\hat{\imath} \hat{\jmath}}[K_{\hat{\imath} \hat{\jmath}}]$ is the 
trace of $[K_{\hat{\imath} \hat{\jmath}}]$ and
$S_{_{\hat{\imath} \hat{\jmath}}} = \text{\textrm{diag}} ( \sigma, 
-\vartheta_{\varphi }, -\vartheta_{z} )$ is the surface stress-energy tensor, 
with $\sigma$ the surface energy density and $\vartheta_{\varphi , z}$ the 
surface tensions. Then replacing Eqs. (\ref{e8a}), (\ref{e8b}) 
and (\ref{e8c}) in Eq. (\ref{e10}) we obtain
\begin{equation}
  \sigma = - \frac{\sqrt{1 + f ( a ) \dot{a}^2}}{8 \pi \sqrt{f ( a )}} \left[
  \frac{g' ( a )}{g ( a )} + \frac{h' ( a )}{h ( a )} \right],
\label{e11}
\end{equation}
\begin{equation}
\vartheta_{\varphi} = - \frac{1}{8 \pi \sqrt{f(a)} \sqrt{1 + f(a)\dot{a}^2}} 
\left\{ 2 f(a) \ddot{a} + f (a) \left[ \frac{h'(a)}{h(a)} + \frac{2 f'(a)}
{f(a)} \right] \dot{a}^2 + \frac{h'(a)}{h (a)} + \frac{f'(a)}{f(a)} \right\},
\label{e12}
\end{equation}
\begin{equation}
\vartheta_{z} = - \frac{1}{8 \pi \sqrt{f ( a )} \sqrt{1 + f(a) \dot{a}^2}}
\left\{ 2 f(a) \ddot{a} + f(a) \left[ \frac{g'(a)}{g(a)} + \frac{2 f'(a)}
{f(a)} \right] \dot{a}^2 + \frac{g'(a)}{g(a)} + \frac{f'(a)}{f(a)} \right\} .
\label{e13}
\end{equation}
The surface energy density is negative, indicating the presence of
\textit{exotic} matter at the throat. The negative signs of the tensions mean 
that they are indeed pressures.\\

It is easy to see that $\vartheta_{\varphi }$, $\vartheta_{z}$ and $\sigma$ 
satisfy the equation
\begin{equation}
\vartheta_{\varphi }-\vartheta_{z}=\frac{g(a)h'(a)-g'(a)h(a)}
{g(a)h'(a)+g'(a)h(a)}\sigma .
\label{e14}
\end{equation}

The static equations are obtained with $\dot{a}=0$ and $\ddot{a}=0$ in Eqs. 
(\ref{e11}), (\ref{e12}) and (\ref{e13}):
\begin{equation}
\sigma = - \frac{1}{8 \pi \sqrt{f ( a )}} \left[
\frac{g' ( a )}{g ( a )} + \frac{h' ( a )}{h ( a )} \right],
\label{e15}
\end{equation}
\begin{equation}
\vartheta_{\varphi} = - \frac{1}{8 \pi \sqrt{f ( a )} } 
\left[\frac{h'(a)}{h(a)}+ \frac{f'(a)}{f(a)} \right],
\label{e16}
\end{equation}
\begin{equation}
\vartheta_{z} = - \frac{1}{8 \pi \sqrt{f ( a )}} 
\left[\frac{g' ( a )}{g ( a )} + \frac{f' ( a )}{f ( a )} \right] .
\label{e17}
\end{equation}
Eqs. (\ref{e16}) and (\ref{e17}) can be recast in the form
\begin{equation}
\vartheta_{\varphi}=\alpha(a)\sigma,
\label{e18}
\end{equation}
\begin{equation}
\vartheta_{z}=\beta(a)\sigma ,
\label{e19}
\end{equation}
with
\begin{equation}
\alpha(a)=\frac{g(a)[f(a)h'(a)+f'(a)h(a)]}{f(a)[g(a)h'(a)+g'(a)h(a)]},
\label{e20}
\end{equation}
\begin{equation}
\beta(a)=\frac{h(a)[f(a)g'(a)+f'(a)g(a)]}{f(a)[g(a)h'(a)+g'(a)h(a)]}.
\label{e21}
\end{equation}
The functions $f$, $g$ and $h$ determine the equations of state 
$\vartheta_{\varphi}(\sigma)$ and $\vartheta_{z}(\sigma)$ of the exotic matter 
on the shell.\\

Let us assume that the equations of state for the dynamic case have the same 
form as in the static one, i.e. that they do not depend on the derivatives of 
$a(\tau)$, so $\vartheta_{\varphi}(\sigma)$ and $\vartheta_{z}(\sigma)$ are 
given by Eqs. (\ref{e18}) and (\ref{e19}), with $\alpha $ and $\beta $ of Eqs. 
(\ref{e20}) and (\ref{e21}). Then, replacing Eqs. (\ref{e11}) and (\ref{e12}) 
in Eq. (\ref{e18}) (or Eqs. (\ref{e11}) and (\ref{e13}) in Eq. (\ref{e19})), 
a simple second order differential equation for $a(\tau )$ is obtained
\begin{equation} 
2f(a)\ddot{a}+f'(a)\dot{a}^{2}=0.
\label{e22}
\end{equation}
It is easy to see that 
\begin{equation} 
\dot{a}(\tau )=\dot{a}(\tau _{0})\sqrt{\frac{f(a(\tau _{0}))}{f(a(\tau))}},
\label{e23}
\end{equation}
satisfies Eq. (\ref{e22}), with $\tau _{0}$ an arbitrary (but fixed) time.
Eq. (\ref{e23}) can be put in the form
\begin{equation} 
\sqrt{f(a)}da=\dot{a}(\tau _{0})\sqrt{f(a(\tau _{0}))}d\tau ,
\label{e24}
\end{equation}
that integrating both sides, it gives
\begin{equation} 
\int^{a(\tau )}_{a(\tau _{0})}\sqrt{f(a)}da=\dot{a}(\tau _{0})
\sqrt{f(a(\tau _{0}))}(\tau - \tau _{0}).
\label{e25}
\end{equation}
The time evolution of the radius of the throat $a(\tau )$ is 
formally obtained by calculating the integral and inverting Eq. (\ref{e25}). 

\section{Vacuum cosmic string wormholes}\label{vacuum}

The most general metric which can be associated to a local vacuum cosmic 
string has the Weyl's form
\begin{equation}
ds^{2}= \left( \frac{r}{r_{0}}\right)^{2d(d-1)}(-dt^{2}+dr^{2})+
r^{2}W_{0}^{2}\left( \frac{r}{r_{0}}\right)^{-2d}d\varphi^{2}+
\left( \frac{r}{r_{0}}\right)^{2d}dz^{2},
\label{v1}
\end{equation}
where $r_{0}$ (a scaling length for the radial coordinate), $W_{0}>0$ and 
$d$ are constants. We take $d<1$, so 
$g_{\varphi \varphi }$ is an increasing function of $r$. Then the surface 
energy density and tensions at the throat are
\begin{equation}
\sigma =-\frac{\left( \frac{a}{r_{0}}\right) ^{-d(d-1)}\sqrt {1
+\left( \frac{a}{r_{0}}\right) ^{2d(d-1)}\dot{a}^{2}}}{4\pi a},
\label{v2}
\end{equation}
\begin{equation}
\vartheta_{\varphi} =-\frac{d^{2}\left( \frac{a}{r_{0}}\right) ^{-d(d-1)}+
\left( \frac{a}{r_{0}}\right) ^{d(d-1)}[a\ddot{a}+d(2d-1)\dot{a}^{2}]}
{4\pi a\sqrt {1+\left( \frac{a}{r_{0}}\right) ^{2d(d-1)}\dot{a}^{2}}},
\label{v3}
\end{equation}
\begin{equation}
\vartheta_{z}=-\frac{(d-1)^{2}\left( \frac{a}{r_{0}}\right) ^{-d(d-1)}+
\left( \frac{a}{r_{0}}\right) ^{d(d-1)}[a\ddot{a}+(d-1)(2d-1)\dot{a}^{2}]}
{4\pi a\sqrt {1+\left( \frac{a}{r_{0}}\right) ^{2d(d-1)}\dot{a}^{2}}}.
\label{v4}
\end{equation}

For the static case we have that $\vartheta_{\varphi}=d^{2}\sigma $ and 
$\vartheta_{z}=(d-1)^{2}\sigma $. Keeping these equations of state for 
the dynamic case and using Eq. (\ref{e25}), we obtain
\begin{equation}
\frac{a(\tau )}{r_{0}}=\left\{ \left[ \frac{a(\tau _{0})}{r_{0}}\right] 
^{d(d-1)+1}+\dot{a}(\tau _{0})\left[ 
\frac{a(\tau _{0})}{r_{0}}\right] ^{d(d-1)}[d(d-1)+1]
\frac{\tau -\tau _{0}}{r_{0}}\right\}^\frac{1}{d(d-1)+1}.
\label{v5}
\end{equation}

As $d(d-1)+1$ is positive for all $d$, from Eq. (\ref{v5}) we see that if the 
initial velocity of the throat 
$\dot{a}(\tau_{0})$ is positive, the radius of the throat increases (without 
bounds) with time, while in the case of negative initial velocity it decreases 
to collapse to $a=0$ in a finite time, and if $\dot{a}(\tau_{0})=0$ the 
throat has constant radius $a(\tau_{0})$ (static solution).\\

When $d\neq 0$ the geometry outside the throat could be interpreted as the one 
corresponding to a \textit{wiggly} or \textit{noisy} cosmic string.  
A special interesting case is the straight cosmic string wormhole, which is 
invariant under boosts in $z$ and corresponds to $d=0$, its metric given by
\begin{equation}
ds^{2}= -dt^{2}+dr^{2}+W_{0}^{2}r^{2}d\varphi^{2}+dz^{2}.
\label{s1}
\end{equation}
This geometry is conical with a deficit angle $\Delta \varphi =2\pi (1-W_{0})$ 
if $0<W_{0}<1$ (surplus angle if $W_{0}>1$). The energy density and tensions 
are\\
\begin{equation}
\sigma =-\frac{\sqrt{1+\dot{a}^{2}}}{4\pi a},
\label{s2}
\end{equation}
\begin{equation}
\vartheta _{\varphi }=-\frac{\ddot{a}}{4\pi \sqrt{1+\dot{a}^{2}}},
\label{s3}
\end{equation}
and
\begin{equation}
\vartheta _{z}=-\frac{1+\dot{a}^{2}+a\ddot{a}}{4\pi a\sqrt{1+\dot{a}^{2}}}.
\label{s4}
\end{equation}
The static solution has $\sigma =\vartheta _{z}=-1/4\pi a$ and 
$\vartheta _{\varphi }=0$; and the energy density per unit length is 
$\mu = 2\pi W_{0}a\sigma =-W_{0}/2$. In the dynamic case, using  Eq. 
(\ref{v5}) with $d=0$, we have
\begin{equation}
a(\tau )=a(\tau _{0})+\dot{a}(\tau _{0})(\tau -\tau _{0}), 
\label{s5}
\end{equation}
so in this case there is a simple linear dependence with time for the radius 
of throat.

\section{Superconducting cosmic string wormholes}\label{supercond}

The exterior metric for a superconducting cosmic string has the form 
\cite{bapi}
\begin{equation}
ds^{2}=\left( \frac{r}{r_{0}}\right) ^{-2m}A^{2}(r)\left[ \left( 
\frac{r}{r_{0}}\right) ^{2m^{2}}(-dt^{2}+dr^{2})+W_{0}^{2}r^{2}d\varphi^{2}
\right] +\left( \frac{r}{r_{0}}\right) ^{2m}\frac{1}{A^{2}(r)}dz^{2},
\label{sc1a}
\end{equation}
where
\begin{equation}
A(r)=\frac{\left( \frac{r}{r_{0}}\right) ^{2m}+k}{1+k},
\label{sc1b}
\end{equation}
with $r_{0}$ (a scaling length for the radial coordinate), $k\ge 0$, $W_{0}>0$ 
and $m$ constants. If we take $-1<m<1$, 
$g_{\varphi \varphi }$ is an increasing function of $r$ for every 
(non-negative) value of $k$. The electric current related with this metric is 
\begin{equation}
I=\pm \frac{mW_{0}}{1+k}\sqrt{k},
\label{sc2}
\end{equation}
and the associated magnetic field strength is given by
\begin{equation}
F_{zr}=-F_{rz}=\frac{\pm 2m}{r}\left( \frac{r}{r_{0}}\right) ^{2m}
\left[ \left( \frac{r}{r_{0}}\right) ^{2m}+k\right] ^{-2}(1+k)\sqrt{k}.
\label{sc3}
\end{equation}
With our cut and paste construction, we obtain in this case a wormhole that 
carries a current $I$ along the throat and a magnetic field outside the 
throat, given by Eqs. (\ref{sc2}) and (\ref{sc3}) respectively. If $k=0$, 
there is no current and the magnetic field is zero, and the Weyl's metric is 
recovered (taking $m=-d$).\\

Using Eqs. (\ref{e11}), (\ref{e12}) and (\ref{e13}), the energy density and 
tensions at the throat are given by
\begin{equation}
\sigma =\frac{-\left( \frac{a}{r_{0}}\right) ^{m(1-m)}
\sqrt{(1+k)^{2}+\left[\left(\frac{a}{r_{0}}\right)^{2m}+k \right] ^{2}
\left(\frac{a}{r_{0}}\right)^{2m(m-1)}\dot{a}^{2}}}
{4\pi a \left[ \left( \frac{a}{r_{0}}\right) ^{2m}+k\right] },
\label{sc4}
\end{equation}
\begin{eqnarray}
\vartheta_{\varphi} &=&\frac{-\left( \frac{a}{r_{0}}\right) ^{m(1-m)}}
{4\pi a \left[ \left( \frac{a}{r_{0}}\right) ^{2m}+k \right]  
\sqrt{(1+k)^{2}+\left[\left(\frac{a}{r_{0}}\right)^{2m}+k \right] ^{2}
\left(\frac{a}{r_{0}}\right)^{2m(m-1)}\dot{a}^{2}}}\times
\nonumber \\
&&\left\{ m^{2}(1+k)^{2}+\left[ \left( \frac{a}{r_{0}}\right) ^{2m}+k 
\right]^{2}\left( \frac{a}{r_{0}}\right) ^{2m(m-1)}a\ddot{a}+\right. 
\nonumber \\
&&\left. m\left(\frac{a}{r_{0}}\right)^{2m^{2}}\left[ k^{2}(2m-1)
\left(\frac{a}{r_{0}}\right)^{-2m}+(2m+1)\left(\frac{a}{r_{0}}\right)^{2m}
+4mk\right]\dot{a}^{2}\right\},
\label{sc5}
\end{eqnarray}
and
\begin{equation}
\vartheta_{z}=\vartheta_{\varphi}+\left[ 1+2m-\frac{4mk}{\left( 
\frac{a}{r_{0}}\right) ^{2m}+k}\right] \sigma .
\label{sc6}
\end{equation}

From the static solution we obtain the equations of state
\begin{equation}
\vartheta_{\varphi} =m^{2}\sigma ,
\label{sc7}
\end{equation}
and
\begin{equation}
\vartheta_{z}=\left[ (1+m)^{2}-
\frac{4mk}{\left( \frac{a}{r_{0}}\right) ^{2m}+k}\right] \sigma.
\label{sc8}
\end{equation}
Following Sec. \ref{cylind} we have that the time evolution of the 
radius of the throat is implicitly given by the equation 
\begin{eqnarray}
\frac{k}{p}\left\{ \left[ \frac{a(\tau )}{r_{0}}\right] ^{p}-
\left[ \frac{a(\tau _{0})}{r_{0}}\right] ^{p}\right\} +
\frac{1}{q}\left\{ \left[ \frac{a(\tau )}{r_{0}}\right] ^{q}-
\left[ \frac{a(\tau _{0})}{r_{0}}\right] ^{q}\right\} &=&
\dot{a}(\tau _{0})\left[ \frac{a(\tau _{0})}{r_{0}}\right] ^{m(m-1)}\times 
\nonumber \\
&&\left\{ k+\left[ \frac{a(\tau _{0})}{r_{0}}\right] ^{2m}\right\} 
\frac{\tau -\tau_{0}}{r_{0}},
\label{sc9}
\end{eqnarray}
with $p\equiv m^{2}-m+1$ and $q\equiv m^{2}+m+1$ positive numbers for all 
values of $m$, and $\tau _{0}$ an arbitrary (fixed) time. In all cases the 
Eq. (\ref{sc9}) can be inverted numerically, and in some cases analytically, 
to obtain $a(\tau )$.\\

The velocity and the acceleration of the throat are, respectively, 
\begin{equation}
\dot{a}(\tau )=\dot{a}(\tau _{0})\left[ \frac{a(\tau _{0})}{a(\tau )}\right]
^{m(m-1)}\frac{\left[ \frac{a(\tau _{0})}{r_{0}}\right]^{2m}+k}
{\left[ \frac{a(\tau )}{r_{0}}\right]^{2m}+k},
\label{sc10}
\end{equation}
and
\begin{equation}
\ddot{a}(\tau )=\frac{-m\dot{a}(\tau _{0})^{2}}{a(\tau )}
\left\{ \frac{\left[ \frac{a(\tau _{0})}{r_{0}}\right]^{2m}+k}
{\left[ \frac{a(\tau )}{r_{0}}\right]^{2m}+k}\right\} ^{2}
\left[ \frac{a(\tau _{0})}{a(\tau )}\right]^{2m(m-1)}
\left\{ k(m-1)+(m+1)\left[ \frac{a(\tau )}{r_{0}}\right]^{2m}\right\}.
\label{sc11}
\end{equation}
It is easy to see that the sign of the velocity is given by the sign of the 
initial velocity $\dot{a}(\tau _{0})$ and the acceleration is always negative. 
As a consequence, if the initial velocity is positive, the throat expands 
forever, with decreasing velocity, whether in the case of negative initial 
velocity it contracts to zero radius with increasing (in modulus) velocity. 
In the case of null initial velocity the radius of the throat remains constant.\\

There are two limiting cases of interest, which correspond to small values of 
the current $I$, where $a(\tau)$ can be approximately given in an explicit 
form. For $k\ll 1$ Eq. (\ref{sc9}) gives
\begin{equation}
\frac{a(\tau )}{r_{0}}\approx \left\{ \left( \frac{a(\tau _{0})}{r_{0}}\right) 
^{q}+\dot{a}(\tau _{0})\left( \frac{a(\tau _{0})}{r_{0}}\right) ^{m(m-1)}
\left[ k+\left( \frac{a(\tau _{0})}{r_{0}}\right) ^{2m}\right]
q\frac{\tau -\tau_{0}}{r_{0}}\right\} ^{1/q},
\label{sc12}
\end{equation}
and if $k\gg 1$ we obtain
\begin{equation}
\frac{a(\tau )}{r_{0}}\approx \left\{ \left( \frac{a(\tau _{0})}{r_{0}}\right) 
^{p}+\dot{a}(\tau _{0})\left( \frac{a(\tau _{0})}{r_{0}}\right) ^{m(m-1)}
\left[ 1+\frac{1}{k}\left( \frac{a(\tau _{0})}{r_{0}}\right) ^{2m}\right]
p\frac{\tau -\tau_{0}}{r_{0}}\right\} ^{1/p}.
\label{sc13}
\end{equation}
In both Eqs. (\ref{sc12}) and (\ref{sc13}), we observe that, as in the case 
of vacuum cosmic string wormholes, the radius of the throat behaves like a 
positive power of $\tau $ for non vanishing initial velocity.

\section{Discussion}\label{discu}

In this paper we have developed a general analysis of the dynamics of 
cylindrical thin-shell wormholes, under a reasonable assumption regarding the 
equations of state that relate the tensions with the surface energy density of 
the \textit{exotic} matter at the throat. The temporal evolution of the 
radius of the throat was obtained for the general case. We applied this 
formalism to cylindrical geometries of interest that appear in the context of 
local cosmic strings. An observer outside the throat would not distinguish 
the geometry from that of the exterior of a local cosmic string. For the 
examples studied, corresponding to vacuum and superconducting cosmic string 
wormholes, we found that the temporal evolution of the 
throat depends mainly on its initial velocity: if it is positive the throat 
expands indefinitely, in the negative case it collapses to null radius in a 
finite time, and when it is zero, the radius of the throat remains 
constant. In these examples, oscillatory solutions are 
not possible. There exists a static solution for each value of the throat 
radius, but these solutions are unstable under perturbations in the velocity, 
i.e. instead of oscillating around or damping towards an equilibrium position, 
they collapse or expand forever if a non zero initial velocity is given. 
Indeed, from Eq. (\ref{e23}), the sign of the initial velocity completely 
determines the sign of the velocity at any time, this being a general feature 
of cylindrical thin-shell wormholes under the hypothesis of this work. It is 
easy to see from Eq. (\ref{e22}) that in the general case the acceleration has 
the opposite sign of the derivative of $g_{rr}=f(r)$, so depending on the 
metric considered, it would accelerate or decelerate the expansion or 
contraction of the throat, but without changing the sign of the velocity, 
which is given by its initial value. 

\section*{Acknowledgments}

This work has been supported by Universidad de Buenos Aires (UBACYT X-143,
E.F.E.) and Universidad de Buenos Aires and CONICET (C.S.). Some calculations 
in this paper were done with the help of the package GRTensorII {\cite{grt}}.

\end{document}